\documentclass[aps,twocolumn,showpacs,amsmath,amstex,amssymb,citeautoscript]{revtex4-2}
\usepackage{physics}
\usepackage{siunitx}
\usepackage[colorlinks,linkcolor=magenta,citecolor=magenta]{hyperref}
\usepackage{bm}
\usepackage{graphicx}

\usepackage{tikz}

\usepackage{xcolor}
\usepackage{lipsum}

\begin{document}

\title{Optical Routing via High Efficiency Composite Acoustic Diffraction}

\author{Yuxiang Zhao$^1$}
\email{yuxiangzhao21@m.fudan.edu.cn}
\author{Jiangyong Hu$^1$}
\author{Ruijuan Liu$^1$}
\author{Ruochen Gao$^2$}
\author{Yiming Li$^1$}
\author{Xiao Zhang$^3$}
\author{Huanfeng Zhu$^{4}$}
\author{Saijun Wu$^1$}
\email{saijunwu@fudan.edu.cn}

\address{$^1$Department of Physics, State Key Laboratory of Surface Physics and Key Laboratory of Micro and Nano Photonic Structures (Ministry of Education), Fudan University, Shanghai 200433, China.
}
\address{$^2$State Key Laboratory of Low Dimensional Quantum Physics, Department of Physics, Tsinghua University, Beijing 100084, China.}
\address{$^3$State Key Laboratory of Quantum Optics and Quantum Optics Devices, Institute of Laser Spectroscopy, Shanxi University, Taiyuan 030006, China.}
\address{$^4$State Key Laboratory of Modern Optical Instrumentation, College of Optical Science and Engineering, Zhejiang University, Hangzhou 310027, China.}


\begin{abstract}
Acousto-optical modulation (AOM) is a powerful and widely used technique for rapidly controlling the frequency, phase, intensity, and direction of light. Based on Bragg diffraction, AOMs typically exhibit moderate diffraction efficiency, often less than 90\% even for collimated inputs. In this work, we demonstrate that this efficiency can be significantly improved using a composite (CP) setup comprising a pair of 4-F-linked AOMs, enabling beamsplitting with fully tunable splitting amplitude and phase. The efficiency enhancement arises from two effects, termed "momentum echo" and "high-order rephasing," which can be simultaneously optimized by adjusting the relative distance between the two AOMs. This method is resource-efficient, does not require ultra-collimation, and maintains control bandwidth. Experimentally, we achieved a diffraction efficiency exceeding 99\% (excluding insertion loss) and a 35 dB single-mode suppression of the 0th-order beam, demonstrating a full-contrast optical router with a switching time of less than 100~nanoseconds. Theoretically, we formulate the dynamics of CP-AOM in terms of multi-mode quantum control and discuss extensions beyond the $N=2$ configuration presented in this work. The substantially enhanced performance of CP-AOMs, coupled with reduced acoustic amplitude requirements, may significantly advance our ability to accurately control light at high speeds with low-loss acousto-optics.

\end{abstract}

\maketitle

\section{Introduction}\label{sec:intro}
From advanced quantum control~\cite{Thom2013,Torrontegui2020a, Bluvstein2024} and optical quantum information processing~\cite{Rudolph2017, Madsen2022, Yu2023} to the development of laser technology itself~\cite{Koke2010, Hansel2017, Wang2021}, 
accurate control of light with high enough speed is essential for enabling research advances at the frontiers. When the required control bandwidth exceeds a megahertz, mechanical control methods~\cite{Zupancic2016,Chen2022c} can be replaced by acousto-optical modulations (AOM)~\cite{Donley2005, Bluvstein2022} or electro-optical modulations (EOM)~\cite{Gould2015, He2020a}. While EOMs, based on electronic response, are ultrafast~\cite{Torrontegui2020a}, the high voltage requirement for efficient operation~\cite{Chiow2007} limits their applications in free space setups. When the control timescale is ten nanoseconds or longer~\cite{Gould2015}, acousto-optical modulators become highly convenient for precisely controlling the phase, frequency, intensity, and propagation direction of light. Today, AOMs are ubiquitously used across fields, with applications ranging from ultrafast pulse shaping to quantum information processing.

The Bragg diffraction underlying AOM is a wave-mechanical effect well studied across fields. In ref.~\cite{Liu2022c} an obvious link is highlighted between the paraxial acousto-optical modulation of light, and the time-domain diffraction of atomic matterwave using standing-wave pulses~\cite{Giltner1995, wang2005,
Wu2005, Chiow2008, Cronin2008}. 
Benefited from time-domain pulse shaping~\cite{Louie2023}, Bragg diffraction techniques are under rapid developments~\cite{Parker2016, Plotkin-Swing2018, Asenbaum2020a, Kirsten-Siemb2023, Louie2023} for achieving exceptional performances, even with uncollimated matterwave~\cite{Louie2023}. In contrast, in traditional AOM, even the first-order Bragg diffraction has a moderate efficiency, $R\sim 90\%$ at best in terms of optical power. In AOM, the acoustic field is designed by patterning the piezo transducer from the side~\cite{Balakshy2021, Wu2023}, not as versatile as time-domain pulse shaping for the matterwave control. 
The moderate efficiency limits the usefulness of the technique in applications where optical losses need to be suppressed, such as for photonic quantum information processing~\cite{Rudolph2017,
Madsen2022,Yu2023} or even just for multi-pass accumulation of AOM frequency-shifts~\cite{Zhou2020}. Furthermore, since at these efficiencies the 0$^{\rm th}$-order has substantial residuals,  traditional AOM can not be used as a router to divide the input beams into the $0^{\rm th}$ and the $1^{\rm st}$ orders on demand.

\begin{figure*}[htbp]
    \centering
    \includegraphics[width=1\linewidth]{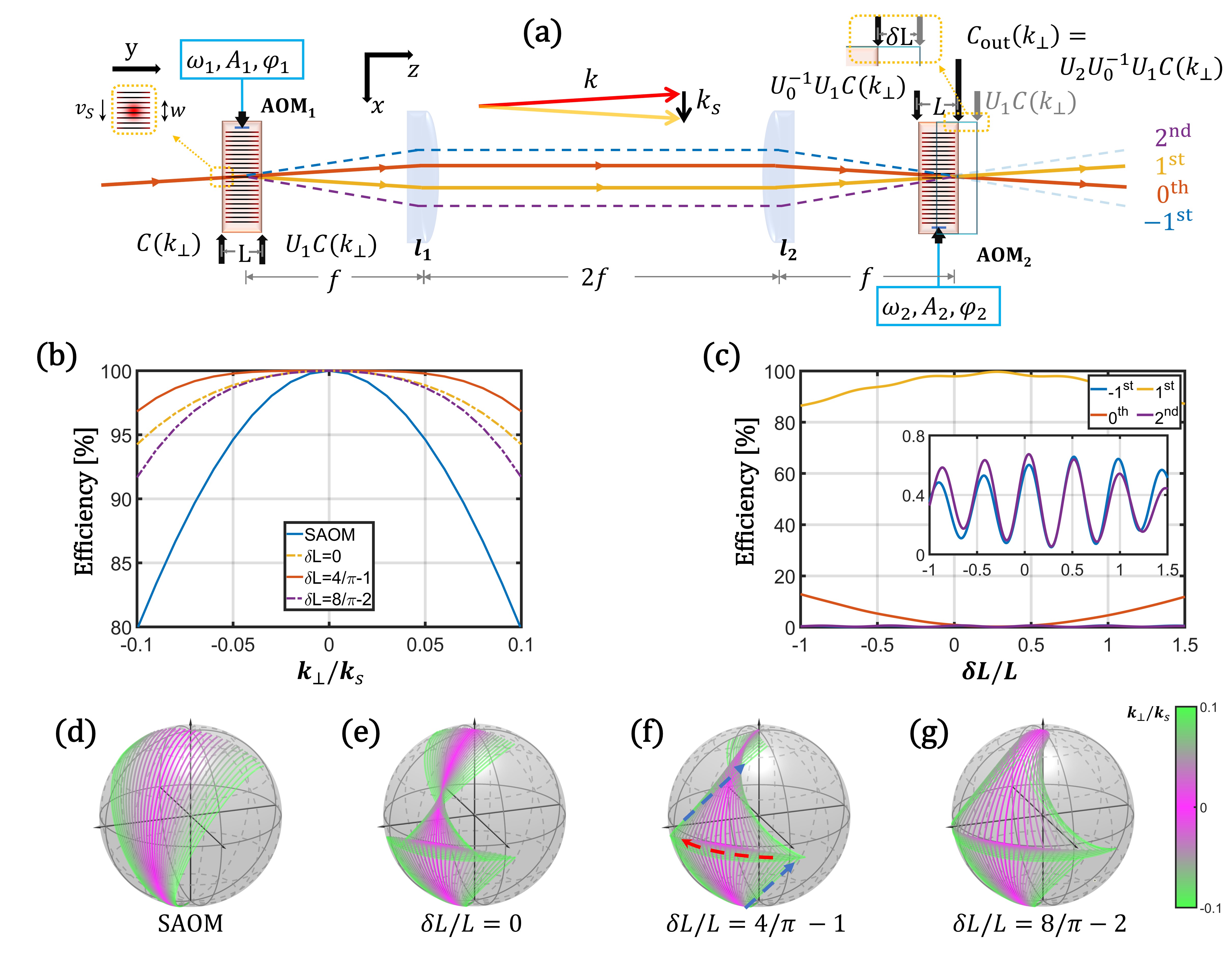}
    \caption{CP-AOM with finely tuned inter-AOM distance $\delta L$ to support ultra-high diffraction efficiency. (a): The setup and the associated wavefront evolution. We consider a collimated input beam from left with width $w$ and wavevector spreading $\Delta k_\perp\approx 1/w\ll k_s$. The wavefront is represented by a $C(k_\perp)$-class to evolve according to Eqs.~\eqref{eq:epsilon_k0}\eqref{eq:C}. The AOM$_2$ is displaced from the perfect imaging plane to AOM$_1$ by $\delta L$. (b): $R(k_\perp)$ vs $k_\perp$ at different $\delta L$ under 2-mode approximation. (c): Numerically simulated optimal diffraction efficiency (Eq.~\eqref{eq:R}) into the $1^{\rm st}$ order as a function of $\delta L$, together with the residuals in the $0^{\rm th}$, $2^{\rm nd}$ and $-1^{\rm st}$ orders. The inset gives the $2^{\rm nd}$ and $-1^{\rm st}$ high-order losses. See Sec.~\ref{sec:num} for details. (d-g): Bloch sphere representation of diffraction dynamics (Eq.~\eqref{eq:Bloch}) for single AOM (d), and for CP-AOM with various $\delta L$ (e-g) as those in (b). The $k_\perp$ is sampled between $-0.1 k_s$ and $0.1 k_s$ as those in (b). In (f) the dashed blue arrows represent $U_{1,2}$ diffractions. The dashed red arrow on the equator represents the $U_0^{-1}$ ``momentum echo'', for the example of $k_\perp=0.1 k_s$ trajectory.
    }
    \label{fig:AOM} 
     \end{figure*}

Inspired by the time-domain matterwave-control perfections~\cite{Wu2005,Louie2023}, one might anticipate exceptional AOM performance by engineering the acoustic field along the light propagation, {\it i.e.} along $z$ in Fig.~\ref{fig:AOM}a within a single AOM~\cite{Balakshy2021}. But the field complexity is likely constrained by acoustic dynamics supportable by finite-sized AOM crystals. Instead, in ref.~\cite{Liu2022c} we propose to precisely image arrays of AOMs to one another to coherently split Bragg-diffraction into an $N$-AOM process, thereby enhancing the controllability in a manner similar to composite-pulse quantum control~\cite{Levitt1986, Low2016}. In ref.~\cite{Liu2022c} the optimal control is discussed in terms of 2-mode dynamics. Experimentally, apart from the exceptional frequency-tuning range,
the optical routing of high-repetition-rate ultrafast pulses, we also found that the achieved $R=95\%$ efficiency was close to the limit set by high-order diffraction losses. Further improving $R$ appeared to require either $N>2$ AOMs, or, by keeping the $N=2$ simplicity, to suppress the high-order losses by increasing the interaction distance $L$ (Fig.~\ref{fig:AOM}a). For the latter case, we were facing a dilemma similar to that for the $N=1$ traditional AOM (Sec.~\ref{sec:simple}): The resonant interaction requires long distance $L$ and large beam-size $w$, necessarily compromises the AOM control bandwidth $\Delta\omega_c\sim v_s/w$ set by the time for the sound wave to trespass the laser beam (Fig.~\ref{fig:AOM}a). In addition, the required large scale, uniform acoustic field could be difficult to generate technically. 

In this work, we demonstrate that AOM efficiency can be perfected through composite diffraction in a resource-efficient manner without sacrificing control bandwidth. The key lies in exploiting wavefront reversal within the CP-AOM in a 4-F imaging setup (Fig.~\ref{fig:AOM}a). By precisely adjusting the position of the second AOM relative to the first (Fig.~\ref{fig:AOM}a), we achieve a remarkable diffraction efficiency of $R=99.2\%$ for a weakly focused laser beam. At this high efficiency, the $0^{\text{th}}$-order output is effectively extinguished (Fig.~\ref{fig:img}c), and single-mode transmission is attenuated by over 35~dB. While our CP-AOM performance is currently limited by imaging aberrations, numerical simulations suggest that an efficiency of $R=99.9\%$ is achievable with improved optical design. The unprecedented performance enables CP-AOM to function as a two-port beamsplitter with fast and fully tunable splitting amplitude and phase. The technique may be highly useful for advancements in laser physics and quantum optics. For example, the exceptional efficiency suggests applications such as efficient iterative frequency shifts to achieve multiple-GHz shifts~\cite{Zhou2020}, or rapid phase-tuning for the coherent combination of multiple CW~\cite{Wang2021} and pulsed~\cite{Klenke2018, Ma2020} lasers. Full-contrast modulation could be useful for tuning the group velocity of mode-locked laser pulses~\cite{Hansel2017}. In quantum optics, using CP-AOM as a fully programmable optical router could enhance high-fidelity, high-speed shaping, and (de)multiplexing~\cite{Liu2022c} of quantum light with AOM devices~\cite{Rudolph2017, Madsen2022, Yu2023}.

We note the diffraction efficiency of $R = 99.2\%$ demonstrated in this work excludes an insertion loss of approximately $5\%$, which can be significantly reduced by employing higher quality optics. To this end, the low RF-power requirements for driving CP-AOM can broaden the selection of acousto-optic materials, such as quartz-based AOMs~\cite{Vidne2003}, which are typically more challenging to drive but offer lower insertion loss.

In the following, we present our work in three sections. First, in Sec.~\ref{sec:principle}, we establish a minimal theoretical model to describe CP-AOM in terms of multi-mode quantum control, and present our numerical results using realistic AOM parameters on optimal diffraction with efficiency approaching $R=99.9\%$. Next, in Sec.~\ref{sec:exp}, we present our experimental results, including the alignment method, and discuss the performance of diffraction-optimized CP-AOM in detail. Finally, in Sec.~\ref{sec:conclusion}, we discuss extension of this work on efficient realization of CP-AOM with single AOM, and on complex optical wavefront engineering with $N>2$ AOMs. We then outline the prospects opened by the high-efficiency CP-AOM technique.

\section{Principles}\label{sec:principle}
Similar to multi-pulse optical control of matter waves~\cite{Wu2005,Hughes2007,Herold2012, Kirsten-Siemb2023, Louie2023}, a paraxially propagating laser beam can be flexibly controlled by multiple acoustic diffractions using RF-driven AOM devices. In both scenarios, with collimated incidence, the diffracted waves often occupy only a few quantized diffraction orders. By truncating the associated Raman-Nath equations (Eq.~\eqref{eq:Norder Raman-Nath}), the diffraction dynamics can be mapped to few-level quantum dynamics. As such, techniques in composite pulse quantum control~\cite{Levitt1986, Low2016} can be utilized for precise and robust wavefront engineering with composite diffractions.

\subsection{Raman-Nath equations}
We first consider AOM$_1$ in the schematic setup in Fig.~\ref{fig:AOM}a. The incident light beam from left is diffracted by the RF-driven acoustic wave along ${\bf e}_x$ inside AOM$_1$. By adjusting the orientation of the AOM, the Bragg-condition is met for efficient transfer of optical energy between the $0^{\rm th}$ and $1^{\rm st}$ orders. 

We consider inside the AOM the index field is given by $n({\bf r},t)=\bar n+\delta n({\bf r},t)$, with
\begin{equation}
	\delta n({\bf r},t)=\eta p \frac{\bar n^2-1}{2\bar n}\cos[ k_s x-\omega_s t+\phi],
	\label{eq:delta_n}
\end{equation}
determined by the acoustic wave propagating along $x$. Here $p$ is the photo-elastic coefficient~\cite{Feldman1978}.  $\omega_s$ is the angular frequency of the sound wave. The $\omega_s=v_s k_s$ relation is fixed by the speed of sound $v_s$ in the AOM crystal. For simplicity we assume a step-function acoustic field, a constant acoustic amplitude $\eta(z)=\eta$ with uniform phase $\phi(z)=\phi$ within the interaction length $0<z<L$. The acoustic amplitude and phase are controlled by $\{A_1,\varphi_1\}$, the amplitude and phase of the RF signal driving AOM$_1$(Fig.~\ref{fig:AOM}a).    

The optical field of the laser beam is written as ${\bf E}({\bf r},t)=\mathcal{E}({\bf r},t) e^{i (k z-\omega t)}+c.c.$. Here $k=\bar n k_0$ and $k_0=\omega/c$. Paraxially, the Helmholtz equation for ${\bf E}$ is reduced to that for the slowly-varying  $\mathcal{E}({\bf r},t)$ envelop,
\begin{equation}
	i(\partial_z+\frac{1}{v_{g,L}}\partial_t) \mathcal{E}=-\frac{1}{2 k}\nabla_{\perp}^2\mathcal{E}-\delta n k_0 \mathcal{E}. \label{eq:2D_SEq}
\end{equation}
Here $v_{g,L}\approx c/\bar n$ is the group velocity for the laser pulse. With the speed of sound to be much smaller than that for light, $v_s\ll v_{g,L}$, in the following analysis we ignore the $\partial_t$-term in Eq.~\eqref{eq:2D_SEq}. We further drop out the $y$-index to specifically consider the diffraction along ${\bf k}_S=k_s{\bf e}_x$. To describe the spatially periodic interaction, we expand $\mathcal{E}(x,z)$ in the Bloch-wave basis for empty lattice, 

\begin{equation}
\mathcal{E}(x,z)=
 \sum_{n} \int_{-k_s/2}^{k_s/2} C_n(k_\perp,z) e^{i((n-1/2) k_s+k_{\perp})x} {\rm d}k_{\perp}.
	\label{eq:epsilon_k0}
\end{equation}
The $C_n(k_{\perp},z)$ is the amplitude of the $n^{\rm th}-$order diffraction with quasi-momentum $k_\perp$. Here we choose to shift the Bloch basis by $\frac{1}{2} k_s$ to ensure the first-order Bragg resonance to meet between the $n=0,1$ orders for $k_\perp=0$. Physically, this requires the incident angle of the collimated laser beam to be tilted by the Bragg angle ${\rm arcsin}(k_s/2 k_0)$ from ${\bf e}_z$ to ${\bf e}_x$. With $K=\eta p \frac{\bar n^2-1}{\bar n} k_0$ to characterize the diffraction strength, the expansion leads to the coupled Raman-Nath equations,
\begin{equation}
    \begin{aligned}
	i\partial_z &C_n(k_\perp,z)=\frac{[(n-\frac{1}{2})k_s+k_\perp]^2}{2 k}C_n(k_\perp,z)\\
    &-\frac{K}{2}e^{i\phi}C_{n-1}(k_\perp,z)-\frac{K}{2}e^{-i\phi}C_{n+1}(k_\perp,z),
	\label{eq:Norder Raman-Nath}
    \end{aligned}
\end{equation}
which can be rewritten  as
\begin{equation}
i\partial_z C= \left(H_0(k_\perp)  +V(K,\phi)\right)C.
    \label{eq:C}
\end{equation} 
The vector $C(k_\perp,z)\equiv (..., C_0, C_1, ...)^{\rm T}$ is referred to as the $C(k_\perp)$-class in the following. The matrix elements $(H_0)_{m n}= \frac{[(n-1/2)k_s+k_\perp]^2}{2 k}\delta_{m,n}$ and $V_{m n}=-\frac{1}{2} (K e^{i\phi}\delta_{m,n+1}+K e^{-i\phi}\delta_{m,n-1})$ are  projections of the first and second terms on the right hand side of Eq.~\eqref{eq:2D_SEq} to the Bloch-wave basis, respectively. We refer $(H_0)_{n n}$ as the propagation constant for the diffraction order $n$.

We formally integrate Eq.~\eqref{eq:C} to obtain the AOM$_1$ propagator for the $C(k_\perp)$-class,
\begin{equation}
U(k_\perp;K_1,\phi_1)=e^{-i (H_0(k_\perp)+V(K_1,\phi_1))L},\label{eq:UAOM}
\end{equation}
which is parametrized by the diffraction strength $K_1$ and acoustic phase $\phi_1$ of the RF-controlled AOM$_1$ (Fig.~\ref{fig:AOM}a). For simplicity we assumed constant $K(z)=K_1$ and $\phi(z)=\phi_1$ according to Eq.~\eqref{eq:delta_n}. Practically the $z-$dependent interaction can be incorporated into the Eq.~\eqref{eq:UAOM} integration in a straightforward manner.

Equation~\eqref{eq:C} conserves the quasi-momentum $k_\perp$. AOM diffraction of any input wavefront $\mathcal{E}(x,z)$ amounts to applying $U(k_\perp;K,\phi)$ in Eq.~\eqref{eq:UAOM} to all the $C(k_\perp)$-class, according to the Eq.~\eqref{eq:epsilon_k0} decomposition. The final wavefront control efficiency are averaged over $k_\perp$ with Eq.~\eqref{eq:R} to be discussed shortly.  As in Fig.~\ref{fig:AOM}a, to facilitate the discussions we use a specific $C(k_\perp)$-class to represent the full wavefront, according to Eq.~\eqref{eq:epsilon_k0}. 

\subsection{Composite diffraction}
We now consider the full Fig.~\ref{fig:AOM}a setup where the AOM$_1$ output is imaged to AOM$_2$ for a second diffraction. We consider the 4-F system images the AOM$_1$ output to $z_2=L+\delta L$, relative to the AOM$_2$ input plane which specifies $z'=0$. Of course, the diffraction orders would already enter AOM$_2$ at $z'=0$ before reaching $z'=z_2$. As being illustrated in Fig.~\ref{fig:AOM}a, the $k_\perp$-class of the input wavefront for AOM$_2$ can therefore be written as $C'=U_0^{-1}  U_1 C(k_\perp)$ using the free-space propagator $U_0=e^{-i H_0 (L+\delta L)}$. With $U_c=U_2 U_0^{-1} U_1$, the AOM$_2$ output can be written as $C_{\rm out}=U_c C(k_\perp)$. With $H_j=H_0(k_\perp)+V(K_j,\phi_j)$, the composite propagator for the $C(k_\perp)$-class is
\begin{equation}
U_c(k_\perp;\{K_j,\phi_j\},\delta L)=e^{-i H_2 L} e^{i H_0 (L+\delta L)}e^{-i H_1 L},\label{eq:Uc}
\end{equation}
parametrized by the RF-controlled $K_j,\phi_j$ and the mechanically adjusted $\delta L$-distance.

\subsection{Mode truncation}\label{sec:mode}
As in Fig~\ref{fig:AOM}a, we now consider collimated Gaussian beam incidence with $C_n(k_\perp) =f(k_\perp)\delta_{n,0}$, with a typical Gaussian profile $f(k_\perp)\propto e^{-k_\perp^2 w^2}$, $w\gg 1/k_s$. Our goal is to optimize the first-order Bragg diffraction efficiency, 
\begin{equation}
R=\int {\rm d}k_\perp R_1(k_\perp) f(k_\perp), \label{eq:R}
\end{equation}
with $R_1(k_\perp)=|(U_c(k_\perp))_{0 1}|^2$ in the CP-AOM parameter space $\{K_1,\phi_1,K_2,\phi_2,\delta L\}$. 

To simplify the analysis, Equation~\eqref{eq:C} can be truncated into a finite set of equations. Clearly, if we decide to only include $n=1-n_0, ..., n_0$ orders, then the diffraction strength must be much smaller than the difference of propagation constants, $K\ll |H_{n_0 n_0}-H_{n_0+1  n_0+1}|\approx 2 n_0 k_r$, so that the coupling between $C_{n_0}$ and $C_{n_0+1}$ is inefficient. The associated high-order loss can be estimated with standard perturbation theory, \begin{equation}
    \mathcal{L}_{n_0}(K) \sim  \overline{|C_{n_0}|^2}\frac{K^2}{16 n_0^2 k_r^2} \label{eq:hLoss}
\end{equation}
in terms of intensity. Here $k_r=k_s^2/2 k$ is a characteristic {\it diffraction constant}. The $\overline{|C_{n_0}|^2}$ factor gives the average $C_{n_0}$ population during AOM diffraction to support the $n_0 \leftrightarrow n_0+1$ coupling loss. Clearly, when $K\ll k_r$, $\overline{|C_{n_0}|^2}$ decreases rapidly with $n_0$.

The simplest truncation of Eq.~\eqref{eq:C} is to only include the $n=0,1$ orders with $n_0=1$ (see Eq.~\eqref{eq:dCh}). By parametrizing $C_0={\rm cos}(\frac{\Theta}{2})e^{-i\Phi/2}$, $C_1={\rm sin}(\frac{\Theta}{2})e^{i\Phi/2}$
, the Bragg diffraction is mapped to 2-level ``spin'' dynamics~\cite{Liu2022c} characterized by a state vector on the Bloch sphere ${\bf n}=({\rm sin}\Theta {\rm cos}\Phi,{\rm sin}\Theta {\rm sin}\Phi,{\rm cos}\Theta)$ obeying
\begin{equation}
\dot{\bf n}={\bf K}\times {\bf n}.\label{eq:Bloch}
\end{equation}
Here we define ${\bf K}=(K{\rm cos}\phi,K{\rm sin}\phi,\Delta)$ as a generalized spatial Rabi-vector. The mismatch $\Delta=k_s k_\perp/ k$ is induced by the quasi-momentum $k_\perp$.

\subsection{Simple AOM: $\mathcal{L}_{1}-$loss and $k_\perp$-broadening}\label{sec:simple}
To understand efficiency constraints in simple AOM diffraction, we first replace the composite $U_c(k_\perp)$ for Eq.~\eqref{eq:R} with the simple $U(k_\perp;K_1,\phi_1)$ in Eq.~\eqref{eq:UAOM}. Full diffraction is designed within the 2-mode approximation by effectively setting $K_1=\theta/L$ with $\theta=\pi$ so that $R(k_\perp=0)=1$. More precisely, for the diffraction of $C(k_\perp)$-class, we have
\begin{equation} 
R(k_\perp)\approx 1-\mathcal{L}_1(K)-a_0 k_\perp^2 L^2.\label{eq:R2level}
\end{equation}
The last term in Eq.~\eqref{eq:R2level} is the leading order correction of the $\theta=\pi-$inversion by the $k_\perp-$mismatch. The coefficient $a_0=(k_s/\pi k)^2$ is derived in Appendix~\ref{sec:derive}. Its effect to the 2-level dynamics is represented by the Bloch sphere dynamics in Fig.~\ref{fig:AOM}d. The diffraction efficiency $R(k_\perp)$ is represented by the solid blue curve in Fig.~\ref{fig:AOM}(b) within the 2-level model. With $K=\pi/L$, the  $\mathcal{L}_1\sim \pi^2/16 k_r^2 L^2$ is the high-order diffraction loss according to Eqs.~\eqref{eq:hLoss}\eqref{eq:dCh}.  

Regarding the choice of the AOM interaction length $L$, Equation~\eqref{eq:R2level} suggests a trade-off between the $\mathcal{L}_1-$loss and $k_\perp$-broadening effects. For  incident laser beam with waist $w$ (Fig.~\ref{fig:AOM}a) so that $\Delta k_\perp\approx 1/w$, an optimal diffraction efficiency with $1- R\sim 1/k_s w$ can be reached with $L=L_{\rm opt}\sim \sqrt{w/k_r}$. To improve the optimal $R$ requires increasing $w$ and $L^2$ together. The wide-$w$ diffraction over long distance $L$ not only requires large-scale AOM fabrication, but also compromises the AOM control bandwidth $\Delta\omega_c=v_s/w$. 

We note that the scaling for $\mathcal{L}_1-$loss and $k_\perp$-broadening in Eq.~\eqref{eq:R2level} are derived assuming the step-function uniform acoustic field between $0<z<L$ (Eq.~\eqref{eq:delta_n}). While smoothing the rising and lowering edges of $\eta(z)$ tends to reduce $\mathcal{L}_1$,  as one expect to occur naturally in AOMs, such perturbations do not change the scaling for the optical control imperfections.

\subsection{Momentum echo}\label{sec:mEcho}
We now consider optimization of Eq.~\eqref{eq:R} with the Fig.~\ref{fig:AOM}a $N=2$ CP-AOM setup. First, as detailed in ref.~\cite{Liu2022c}, for CP-AOM it suffices to set $K_1=K_2=\theta/L$ with $\theta=\pi/2$. By adjusting $\phi_1=\phi_2$, full inversion can be achieved at $k_\perp=0$ within the 2-mode approximation. With $K=\pi/2 L$, high-order diffraction loss for single AOM is suppressed by a factor of four, comparing to $\pi-$inversion, according to Eq.~\eqref{eq:hLoss}. With $R(k_\perp=0)=1$ optimized within the 2-mode approximation, we found that similar to Eq.~\eqref{eq:R2level}, the efficiency for the composite diffraction of $C(k_\perp)$-class can be written as
\begin{equation} 
R(k_\perp)\approx 1-\mathcal{L}_c(K;\delta L)-a(\delta L) k_\perp^2 L^2-b(\delta L) k_\perp^4 L^4.\label{eq:R2levelb}
\end{equation}
The $\mathcal{L}_c(K;\delta L)$ term describes high-order diffraction loss as to be discussed shortly. For the $k_\perp-$broadening, here we include the 2nd and 4th-order $k_\perp L$ corrections. The coefficients $a(\delta L), b(\delta L)$ depend on the CP-AOM displacement $\delta L$ (Fig.~\ref{fig:AOM}a). The expressions are given in Appendix~\ref{sec:derive} where we further show that $a(\delta L)$ is nullified at $\delta L_{\rm opt}=(4/\pi-1)L$ where $b=(4-\pi)^2 (k_s/ \pi k)^4$.
As such, the second order $k_\perp-$broadening is suppressed.  In Fig.~\ref{fig:AOM}b we plot $R(k_\perp)$ as a function of $k_\perp$. The Bloch sphere dynamics of the state vector ${\bf n}$, according to Eq.~\eqref{eq:Bloch}, are given in Fig.~\ref{fig:AOM}(d-g) for different choices of $\delta L$. The enhanced tolerance to $k_\perp-$broadening at the optimal $\delta L_{\rm opt}$ can be exploited to support efficient Bragg diffraction of small-$w$ beams with accordingly enhanced control bandwidth.

The rephasing effect generated by \( U_0^{-1} = e^{i H_0 (L + \delta L)} \) is analogous to the spin echo used in nuclear magnetic resonance (NMR) to suppress inhomogeneous broadening~\cite{Levitt1986}. As illustrated by the red dashed arrow in Fig.~\ref{fig:AOM}(f), the mismatch \(\Delta = k_s k_\perp / k\) causes the state vector \(\mathbf{n}\) with different $k_\perp$ to spread during \( U_1 \). The \( U_0^{-1} \) operation, achieved through 4-F imaging, then reverses this spreading near the equator, setting the stage for the \( U_2 \) diffraction to complete the inversion. With a suitable choice of $\delta L$, the $U_0^{-1}$ reversal is optimally deep to ensure a nearly perfect inversion by $U_2$. We borrow the NMR concept to name the effect as a ``momentum echo''. 


\subsection{High-order rephasing}\label{sec:highOrder}

We now consider the $\mathcal{L}_c(K;\delta L)$ term in Eq.~\eqref{eq:R2levelb}. For $K\ll k_r$ and with the $n=0,1$ truncation to Eq.~\eqref{eq:C}, the $\mathcal{L}_1$ loss is primarily due to the off-resonant $C_{-1},C_2-$couplings. We evaluate the coupling loss by expanding the basis to $n=-1,0,1,2$ orders into Eq.~\eqref{eq:C}. The resulting 4-mode coupling equations 
can be simplified with a change of variables, $C_l^{\pm}=C_0\pm C_1 e^{i\phi}$, $C_h^{\pm}=C_{-1}e^{-i\phi}\pm C_2 e^{2i\phi}$, for the $k_\perp=0$ class, since both the $V(K,\phi)-$coupling and the free $H_0(0)$ in Eq.~\eqref{eq:UAOM} conserve the $\pm$sign symmetry. We focus on the $k_\perp=0$ class in this section.  Generally, with $|C_h^{\pm}|_{\rm in}\ll |C_l^{\pm}|_{\rm in}$ at the AOM incidence, the $V(K,\varphi)$ coupling leads to small shifts of $C_h^{\pm}$ amplitudes, $\delta C_h^{\pm}$, that can be perturbatively evaluated to the leading $K/ k_r $ order as 
\begin{equation}
\delta C_h^{\pm}\approx \frac{K(1-e^{i 2 k_r L})}{4 k_r}(C_l^{\pm})_{\rm in},\label{eq:dCh}
\end{equation}
Again, here we assumed the step-function acoustic field specified in Eq.~\eqref{eq:delta_n}. The $\mathcal{L}_1(K)\sim |\delta C_h^{\pm}|^2$ is just the $n_0=1$ special example of Eq.~\eqref{eq:hLoss}. Practically the edge-smoothed acoustic field $\eta(z)$ (Eq.~\eqref{eq:delta_n}) helps to reduce $\mathcal{L}_c$, but the $K/k_r$ scaling is robust.  

Clearly, the $2k_r$ difference in propagation constants in $H_0$ suppresses the couplings between $C_l$ and $C_h$. In addition, for the $k_\perp=0$ class, we have 
\begin{equation}
    \begin{aligned}    
    C_l^{\pm}(z)&=C_l^{\pm}(0)e^{-i \frac{1}{4}k_r z},\\
    C_h^{\pm}(z)&=C_h^{\pm}(0)e^{-i \frac{9}{4} k_r z}\label{eq:f}
    \end{aligned}
\end{equation}
during free propagation.  Equations~\eqref{eq:dCh}\eqref{eq:f} suggest the high-order loss for the CP-AOM setup takes the form 
\begin{equation}
    \mathcal{L}_c(K;\delta L)=\mathcal{L}_1(K)|\mathcal{A}_1 e^{i 2 k_r (L+\delta L)} + \mathcal{A}_2|^2.\label{eq:hLossC}
\end{equation}
Here $\mathcal{A}_{1,2}=O(1)$ are complex coefficients that depend on detailed diffraction dynamics in specific AOMs. 

We refer the suppression of $\mathcal{L}_c(K;\delta L)$ in Eq.~\eqref{eq:R2levelb} at specific $\delta L$, according to Eq.~\eqref{eq:hLossC}, as a high-order rephasing effect.  As illustrated in Fig.~\ref{fig:AOM}c, the suppression is achieved with a periodic array of optimal $\delta L$, which are decided by the phase factor $\phi_r=2k_r(L+\delta L)$ as well as the phase angle of $\mathcal{A}_1/\mathcal{A}_2$. Of course, we hope the $(\mathcal{A}_1 e^{i 2 k_r (L+\delta L)} + \mathcal{A}_2)$ factor to be minimized at the $\delta L_{\rm opt}$ associated with Eq.~\eqref{eq:R2levelb}, so that the high-order rephasing and the momentum echo described in the last section are optimized simultaneously (see the Fig.~\ref{fig:AOM}c example). Practically,  with $L$ itself fixed, the simultaneous optimization of the momentum echo and high-order rephasing effects can be achieved by finely adjusting $k_s=\omega_s/v_s$ and thus the diffraction constant $k_r$.

\subsection{Numerical simulation}\label{sec:num}
In Fig.~\ref{fig:AOM}c the momentum echo and high-order rephasing effects are illustrated with numerical simulation of the full Eq.~\eqref{eq:Norder Raman-Nath} that models the CP-AOM diffraction of a Gaussian beam at $\lambda=780~$nm. The parameters in the simulation, $w=185~\rm{\mu m}$, $\omega_s=2\pi \cross 80~\rm{MHz}$, $v_s=4260~\rm{m/s}$, $\bar n=2.23$ for AOM based on TeO$_2$ crystal, are chosen according to the experimental setup to be introduced next. With the diffraction strength $K=\pi/2L\approx 0.09~{\rm mm}^{-1}$,  $K\ll k_r$ is well satisfied for $k_r=k_s^2/2k \approx 0.9~{\rm mm}^{-1}$. Here, we choose the interaction length $L=17~$mm for the square-shaped acoustic field. The value is close to the experimental estimation. In addition, it ensures $(\mathcal{A}_1 e^{i 2 k_r (L+\delta L)} + \mathcal{A}_2)$ is minimized at $\delta L_{\rm opt}$. We emphasize that the same fine-tuning can be achieved with fixed $L$ but varying $\omega_s=v_s k_s$, as suggested in the last section. 

First, to illustrate the momentum echo, we compute the $k_\perp$-dependent diffraction efficiency $R(k_\perp)=|\left(C_{\rm out}\right)_1(k_\perp)|^2$ under two-mode approximation, {\it i.e.},  with the $n=0,1$ truncation to Eq.~\eqref{eq:Norder Raman-Nath} (Sec.~\ref{sec:mode}). The initial condition is $C_n=\delta_{n,0}$. The results in Fig.~\ref{fig:AOM}b are obtained by integrating Eq.~\eqref{eq:Norder Raman-Nath} for $U_1 C(k_\perp)$ (blue line, Eq.~\eqref{eq:UAOM}) and $U_c C(k_\perp)$ (other lines, Eq.~\eqref{eq:Uc}). The advantage of choosing $\delta L=\delta L_{\rm opt}$ is clearly seen with the red curve in Fig.~\ref{fig:AOM}b, where a nearly ideal $R(k_\perp)$ is maintained under the two-mode approximation for $|k_\perp|<0.05 k_s$. 

Next, we numerically integrate the full Eq.~\eqref{eq:Norder Raman-Nath} for the Gaussian beam input, via Eq.~\eqref{eq:2D_SEqb} as detailed in Appendix~\ref{app:num}, to obtain the $k_\perp-$averaged diffraction efficiency $R$ according to Eq.~\eqref{eq:R}. Losses to $n=-1,2$ orders and the $n=0$ residual are also recorded. The oscillatory $R_{-1,2}$ as a function of $\delta L$ is with $2 k_r$ frequency, according to Eq.~\eqref{eq:hLossC}. At $\delta L=\delta L_{\rm opt}$, the $R_{-1,2}$ loss and the $R_0$ residual are simultaneously suppressed, leading to an optimal $R_{\rm opt} = 99.7\%$. The losses to $n=-1,2$ orders are blow $0.1\%$. The remaining $0.2\%$ loss is primarily in the $C_0$ residual, as the momentum echo is not able to completely suppress the $k_\perp$-mismatch (Fig.~\ref{fig:AOM}f). By increasing $w$ to $400~\mu$m to reduce the incident $k_\perp$-spreading, $R_{\rm opt}=99.9\%$ is reached in the simulation. We performed additional simulations to verify that the simulated performance is robustly achievable with realistic AOM parameters and {\it improves further} for edge-smoothed acoustic amplitude $\eta(z)$. 

Finally, without resorting to detailed derivations, in Appendix~\ref{sec:hLoss} we show numerically that $\mathcal{L}_c(K,\delta L)$, with $\delta L_{\rm opt}$ optimized at $K=\pi/2 L$, decreases monotonically with $K$ for $K\in [0,\pi/2L]$. The momentum echo is hardly affected by the reduced $K$ either. Therefore, we expect nearly ideal performance of CP-AOM as a two-port beamsplitter with fully programmable diffraction efficiency and phase~\cite{Liu2022c}.

\section{Experimental demonstration}\label{sec:exp}

\begin{figure}[htbp]
    \centering
    \includegraphics[width=1\linewidth]{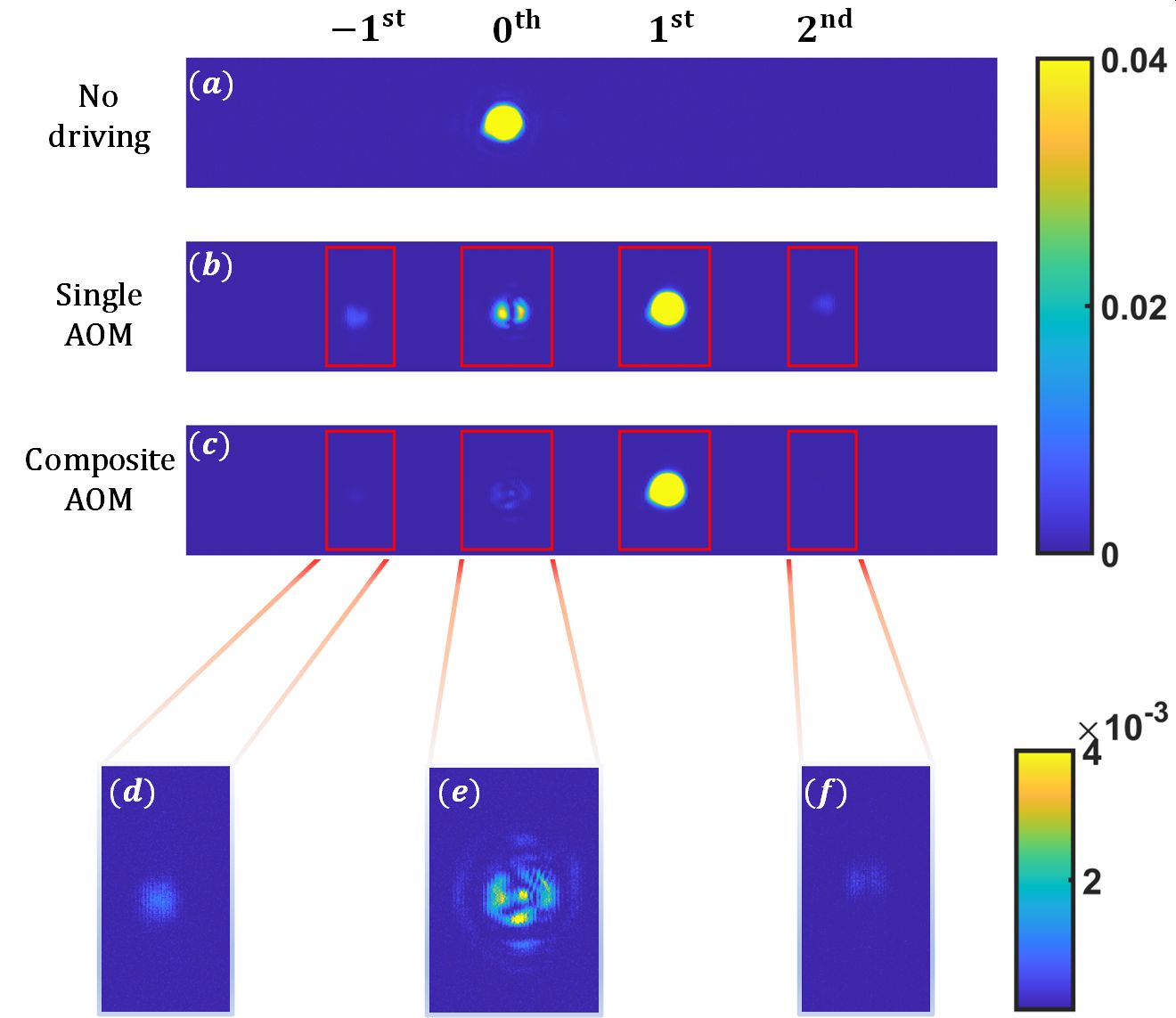}
    \caption{Intensity profiles of the CP-AOM diffraction output, normalized by the peak intensity. (a): The $0^{\rm th}$-order output with AOM$_{1,2}$ off. (b): Output of optimal diffraction by AOM$_{1}$.  (c-f):  Output of optimal diffraction by CP-AOM. 
    }
    \label{fig:img} 
     \end{figure}
\subsection{Setup and measurement procedure}

The schematic of the CP-AOM setup is outlined in Fig~\ref{fig:AOM}a. We use commercially available AOMs (G$\&$H 3080-122) operating around the designed $\omega_s=2\pi\times 80~$MHz frequency, driven by phase-coherent RF-signals from a multi-channel direct-digital synthesizer (Novatech~409B). The optical wavelength is $\lambda=780~$nm. The Gaussian beam input for AOM$_1$ is collimated with a waist size of $w=185~\mu$m ($1/e^2$ intensity). The AOM$_1$ output is imaged by a 4-F system to AOM$_2$. We choose a pair of $f=100~$mm achromatic lenses (Thorlabs AC254-100-B) to ensure aberration-free 4-F links of the $n=-1,0,1,2$ diffraction orders by AOM$_{1,2}$. 

The CP-AOM output is detected by a scientific grade CMOS camera  (Thorlabs CS165MU/M). The four output diffraction orders of interest, referred to as $I_n, n=-1,0,1,2$ in Fig.~\ref{fig:AOM}a, are recorded by four areas on the camera sensor (Fig.~\ref{fig:img})  denoted as $S_n$ in the following. Higher-order diffractions are also recorded. By substantially increasing the input laser power, we verified that the higher diffraction orders contribute negligible when CP-AOM operation is close to be optimal, and can hardly be detected ($<0.01\%$ in power). These orders are therefore ignored in our following analysis. To retrieve Bragg-diffraction power efficiency $R$ (Eq.~\eqref{eq:R}), twenty $I_n$ measurements interleaved with background measurements are repeated and averaged. Typical background-subtracted and averaged images are displayed in Fig.~\ref{fig:img}. We obtain optical power $P_n=\sum_{S_n} I_n$ within each $S_n$. The Bragg diffraction power efficiency is then evaluated as $R=P_1/\sum P_j$.

Clearly, accurate measurement of the diffraction efficiency $R$ in our setup relies on a uniform and linear camera sensor response. By simultaneously measuring the laser power split into two paths, one with $P_0$  (AOMs off) by the camera and one by a calibrated power-meter (Thorlabs S150C), we verify that the camera sensor readout is highly linear when the incident laser power varies by 25 dB. Furthermore, to account for non-uniform camera sensitivity, we translate the motorized-stage mounted camera sensor along $x$ and repeatedly measure $P_0$ (AOMs off) with different $S_n$. The relative responsivity $\xi_n$ is found to deviate by up to $20\%$ across the chip, primarily due to the sensor etalon effect.  We apply the $\xi_n$ correction to all the $P_n$ measurements in the following. As the etalon effect is stable in our temperature-stabilized lab,  we estimate this $\xi_n$ correction to be accurate within $2\%$ to impact negligibly the uncertainty of our $1-R$ measurement in the following.

Finally, in addition to camera measurements, we use a flip mirror to couple the $n=0,1$ output orders into single-mode fibers when necessary. In static measurements, the single-mode outputs are monitored by the S150C power-meter. In transient measurements, we use a fast detector (Thorlabs PDA8GS) to evaluate the switching time of the CP-AOM.

\subsection{Co-propagating geometry}\label{sec:geometry}
In Fig.~\ref{fig:AOM}a, the diffractions by the 4-F-linked AOM$_{1,2}$ share the same order number $n$ with identical $\Delta\omega_n=n\omega_s$ frequency shifts. Consequently, the interference by the overlapping AOM$_{1,2}$ orders are phase-stable. We refer this CP-AOM configuration as being with "co-propagating" acoustic fields.

The pairing of identical orders in Fig.~\ref{fig:AOM}a can be modified by reversing the direction of acoustic-wave propagation in AOM$_2$, as in our previous work~\cite{Liu2022c}. From Eq.~\eqref{eq:delta_n}, the $x\rightarrow -x$ reversal is equivalent to the $\omega_s$ sign reversal. The overlapping diffraction by AOM$_{1,2}$ therefore contains $\pm \omega_s$ sidebands. In our previous work~\cite{Liu2022c}, the resulting $2\omega_s$ interference is exploited for routing a synchronized mode-locked laser into the $n=0, 1$ output directions. We refer that setup as CP-AOM in the ``counter-propagating'' geometry.

Evidently, with the light speed $v_{g,L}=c/n$ in Eq.~\eqref{eq:2D_SEq} effectively set to infinity, the diffraction dynamics are determined by the instantaneous phases $\phi_j(t)=\phi_j(0)\pm \omega_s t, \, j=1,2$ for AOM$_{1,2}$, in both the co-propagating and counter-propagating geometries. In this work, we exploit the co-propagating geometry to demonstrate exceptional diffraction efficiency with a CW-laser input. Our conclusions are directly applicable to the counter-propagating geometry, enabling synchronized control of mode-locked laser pulses by sampling specific instances of $\phi_j(t)$~\cite{Liu2022c}.

\subsection{Optimizing CP-AOM}

The procedure for optimizing the CP-AOM diffraction is similar to that in our previous work~\cite{Liu2022c}. Within the 2-mode approximation, AOM$_{1,2}$ serve as the first and second beamsplitters of a Mach-Zehnder interferometer, respectively. The optimization process is thus aimed at maximizing the interferometric contrast. The observables during optimization are the $n=-1, 0, 1, 2$ output beams on the camera (Fig.~\ref{fig:img}). In this work, to enhance CP-AOM performance, particular attention is given to fine-tuning $\delta L$, the relative displacement between the AOM$_2$ output and the (imaged) AOM$_1$ output (Fig.~\ref{fig:AOM}a). The procedure is as follows.

First, AOM$_{1,2}$ are switched on one by one. The diffraction strength for each AOM is adjusted with the RF-amplitudes $A_{1,2}$ to ensure moderate efficiency $R_{1,2} \leq 50\%$, {\it i.e.}, with intensities $I_1 \leq I_0$ observed on the camera. Then, by adjusting the orientation of AOM$_2$ relative to AOM$_1$, the $n=1$ diffractions from AOM$_{1,2}$ overlap as closely as possible on the camera.

Next, AOM$_{1,2}$ are turned on simultaneously to enable diffraction interference. The RF frequency $\omega_1$ is detuned from $\omega_2$ by a small value, {\it e.g.}, $\delta \omega = 2\pi \times 0.5$ Hz, to scan the relative phase $\phi_{12} = \phi_1 - \phi_2$. Due to the interference, the linearly increasing $\phi_{12}$ leads to oscillatory $I_{0,1}$ output on the camera with $2\pi/\delta \omega$ periodicity. While the first-order diffractions from AOM$_{1,2}$ overlap closely after the last step, any mode mismatch results in temporally varying $I_{1,2}$ distributions. To improve the mode-matching, the 3D location as well as the orientation of AOM$_{1,2}$ are finely adjusted. This fine alignment results in uniformly interfering $I_{1,2}$ on the camera.

Finally, the RF amplitudes $A_{1,2}$ are finely adjusted to maximize the peak $R \propto P_1 / \sum P_j$ during the $\delta \omega$ frequency scan. The peak $R$ is recorded. The optimization is repeated as a function of the relative distance $\delta L$ and then around the designed frequency $\omega_s$ if necessary. As such, the optimal $R_{\rm opt}$ is reached at a specific $\delta L_{\rm opt}$, as expected from Fig.~\ref{fig:AOM}c.

\subsection{Results} 

\subsubsection{Optimal diffraction}
A typical output of the optimal CP-AOM diffraction is shown in Fig.~\ref{fig:img}c. Compared to the AOM-off case in Fig.~\ref{fig:img}a, the 0$^{\rm th}$ order appears simply shifted to the 1$^{\rm st}$ order, despite the display being strongly saturated with the colorbar scaled to 4\% of the peak $I_n$ value. Closer inspection (Fig.~\ref{fig:img}(d-f)) indicates that the residual optical energy is split into $n= -1, 0, 2$ orders, with the dominant contribution from the irregularly-shaped $I_0$ residuals. 

From Fig.~\ref{fig:img}c, a Bragg-diffraction efficiency of $R=99.26\%$ is determined by analyzing $I_n$ for $n=-1, 0, 1, 2$. Considering potential losses due to higher-order diffraction, the peak efficiency is estimated to be $R=99.2\%\pm 0.1\%$. 
The optimal CP-AOM operation is remarkably stable over hours, due to common-mode rejection inherent to the 4-F setup against the vibrational and drifting noises. While the peak efficiency may degrade slightly over long time, it can be restored electronically, simply by adjusting the relative phase $\varphi_{12}$ of the RF signals.

The exceptional CP-AOM performance is due to the momentum echo technique that suppress the $I_0$ residual, and the high-order rephasing technique that suppresses the $I_{-1,2}$ losses. To highlight the difference made by
CP-AOM, in Fig.~\ref{fig:img}b we provide the intensity image profiles for typical single-AOM diffraction. One clearly see much stronger losses in the $n=-1,0,2$ orders. Comparing to Fig~\ref{fig:img}e, the stronger $I_0$ residual is also simpler in shape. As discussed in Sec.~\ref{sec:mEcho}, the $k_\perp-$mismatch leads to inefficient AOM diffraction. Conversely, efficient diffraction of sufficiently small $k_\perp$ components serves as an effective mode-filtering process. Comparing to CP-AOM assisted by the momentum echo (Fig.~\ref{fig:AOM}f), efficient single-AOM diffraction can only occur within a much narrower $k_\perp$ distribution  (Fig.~\ref{fig:AOM}d).

Nevertheless, the $R=99.2\%$ CP-AOM diffraction efficiency is still limited by the $I_0$ residuals as in Fig.~\ref{fig:img}e. The irregular $I_0$ suggests that the inefficient diffraction is caused by aberrations of 4-F imaging system in this work. The aberrations reduce the overlap between the AOM$_{1,2}$ diffractions and therefore degrade the wavefront reversal (Eq.~\eqref{eq:Uc}). With more dedicated imaging optics, we expect further enhanced CP-AOM diffraction toward $R=99.9\%$ as suggested by numerical simulations.
\subsubsection{Frequency scan}
\begin{figure}[htbp]
    \centering
    \includegraphics[width=1\linewidth]{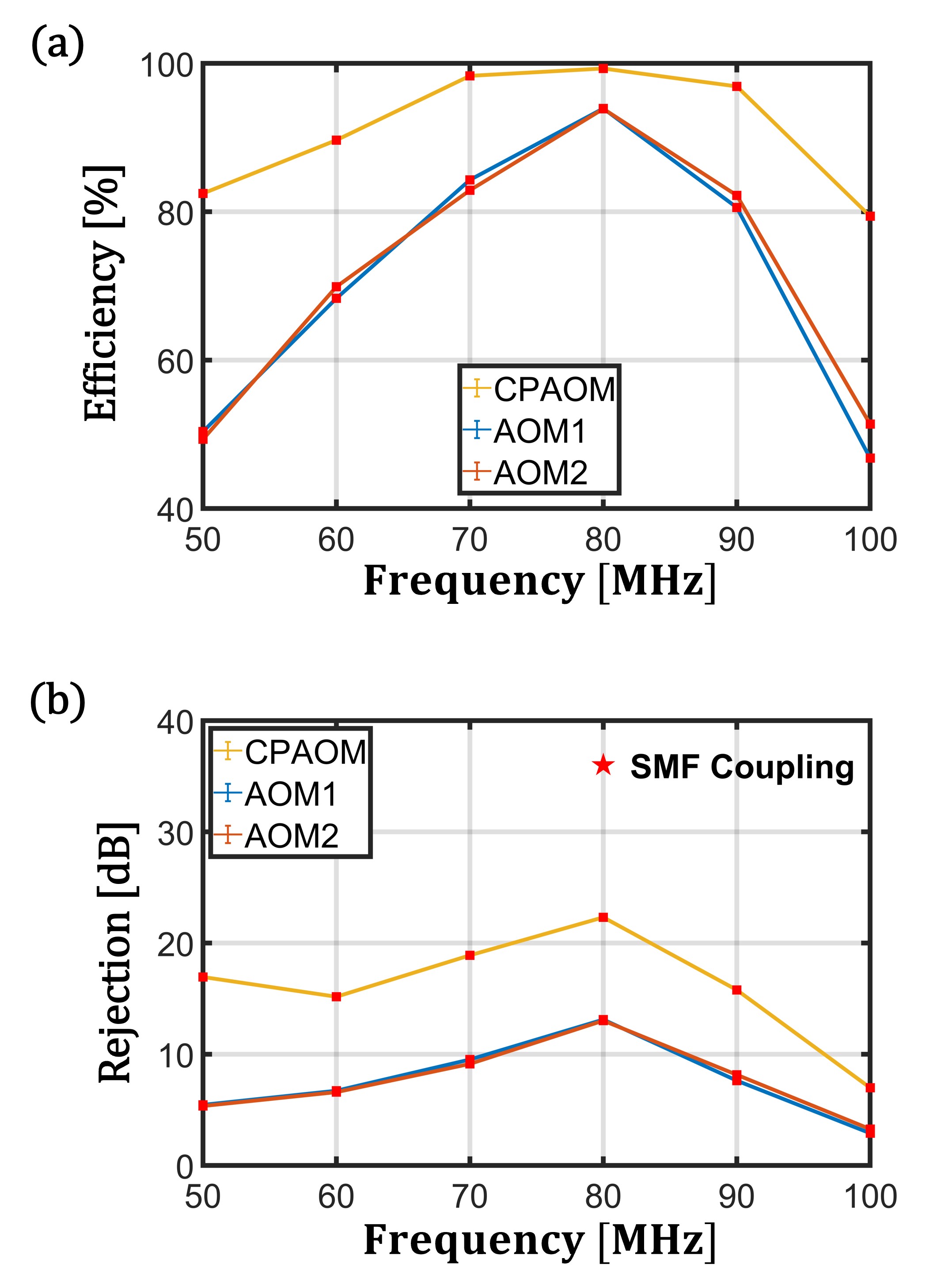}
    \caption{Diffraction efficiency $R$ (a) and $I_0$ rejection ratio $\alpha$ (b) as a function of the operation frequency $\omega_s$ of the CP-AOM aligned at 80~MHz. In (b) a star symbol suggests $\alpha_s$ obtained with single-mode fiber coupling (Sec.~\ref{sec:single}).   
    }
    \label{fig:scan} 
 \end{figure}
 As detailed in Ref.~\cite{Liu2022c}, CP-AOM enables efficient diffraction beyond the Bragg condition through quasi-phase-matching, where the acoustic-optical phase mismatches of AOM$_{1,2}$ cancel each other. Here, after optimizing CP-AOM at the designed frequency of $\omega_s=2\pi\times 80$~MHz, we shift $\omega_{S}$ to other values and optimize $R(\omega_s)$ electronically by adjusting the RF amplitudes $A_{1,2}$ and relative phase $\varphi_{12}=\varphi_1-\varphi_2$.  The optimal $\{A_{j,{\rm opt}}(\omega_s), \varphi_{j,{\rm opt}}(\omega_s)\}$ are saved as a parameter map, allowing us to select the carrier frequency $\omega_s$ of interest for driving CP-AOM while maintaining high diffraction efficiency $R(\omega_s)$.



The CP-AOM capacity for wideband operation is illustrated in Fig.~\ref{fig:scan}a, which shows the RF-optimized efficiency $R(\omega_s)$ during a $\omega_s$ scan from 50~MHz to 100~MHz. For comparison, results from the same $\omega_s$ scan for single AOM$_{1,2}$ are provided. Remarkably, quasi-phase-matching enables CP-AOM to achieve efficiency beyond 90\% over a range of more than 30~MHz. As discussed in ref.~\cite{Liu2022c}, wideband modulation is highly valuable for CP-AOM to rapidly shift the phase, frequency~\cite{Koke2010}, and output direction~\cite{Bluvstein2022} of the first-order diffraction.


Efficient 1$^{\rm st}$-order diffraction is accompanied by effective extinction of the incident 0$^{\rm th}$-order beam. In Fig.~\ref{fig:scan}b, we plot the $I_0$ rejection ratio, $\alpha = 10 \log(P_{0,\rm on}/P_{0,\rm off})$. Here, $P_{0,\rm off/on}$ are the total power of $I_0$ recorded on the camera within $S_0$ (Fig.~\ref{fig:img}), measured when the optimal AOM operation is off and on, respectively. Similar to Fig.~\ref{fig:scan}a, the attenuation data for the optimal CP-AOM and single-AOM are shown in Fig.~\ref{fig:scan}b. The CP-AOM achieves a remarkable $\alpha = 22$~dB attenuation at $\omega_s=2\pi\times 80$~MHz, demonstrating its capability as a high-contrast two-port splitter in free space.


\subsubsection{Single-mode attenuation}\label{sec:single}
The $I_0$ residuals shown in Fig.~\ref{fig:img}(c,e) result from spatial-mode filtering by CP-AOM, displaying an irregular wavefront profile. By coupling the $I_0$ beam into a single-mode fiber, we expect to enhance the $I_0$-rejection ratio further.

We couple both $I_{0,1}$ into single-mode fibers with coupling efficiencies of about $80\%$ for both ports. When CP-AOM is switched on, the $I_0$ attenuation is found to be enhanced from the free-space value to $\alpha_{\rm s}\approx 36~$dB. For comparison, the $I_1$ single-mode attenuation of $\sim 50$~dB is obtained when AOMs are switched off. Clearly, the single-mode $I_0$ rejection, which relies on the efficient Bragg-diffraction to remove the energy, is still limited by the $I_0$ residuals as those displayed in Fig.~\ref{fig:img}e. Nevertheless, the $\alpha_{\rm s}>35~$dB-level attenuation is typically sufficient in various applications.

\subsubsection{Full contrast optical routing}
\begin{figure}[htbp]
    \centering
    \includegraphics[width=1\linewidth]{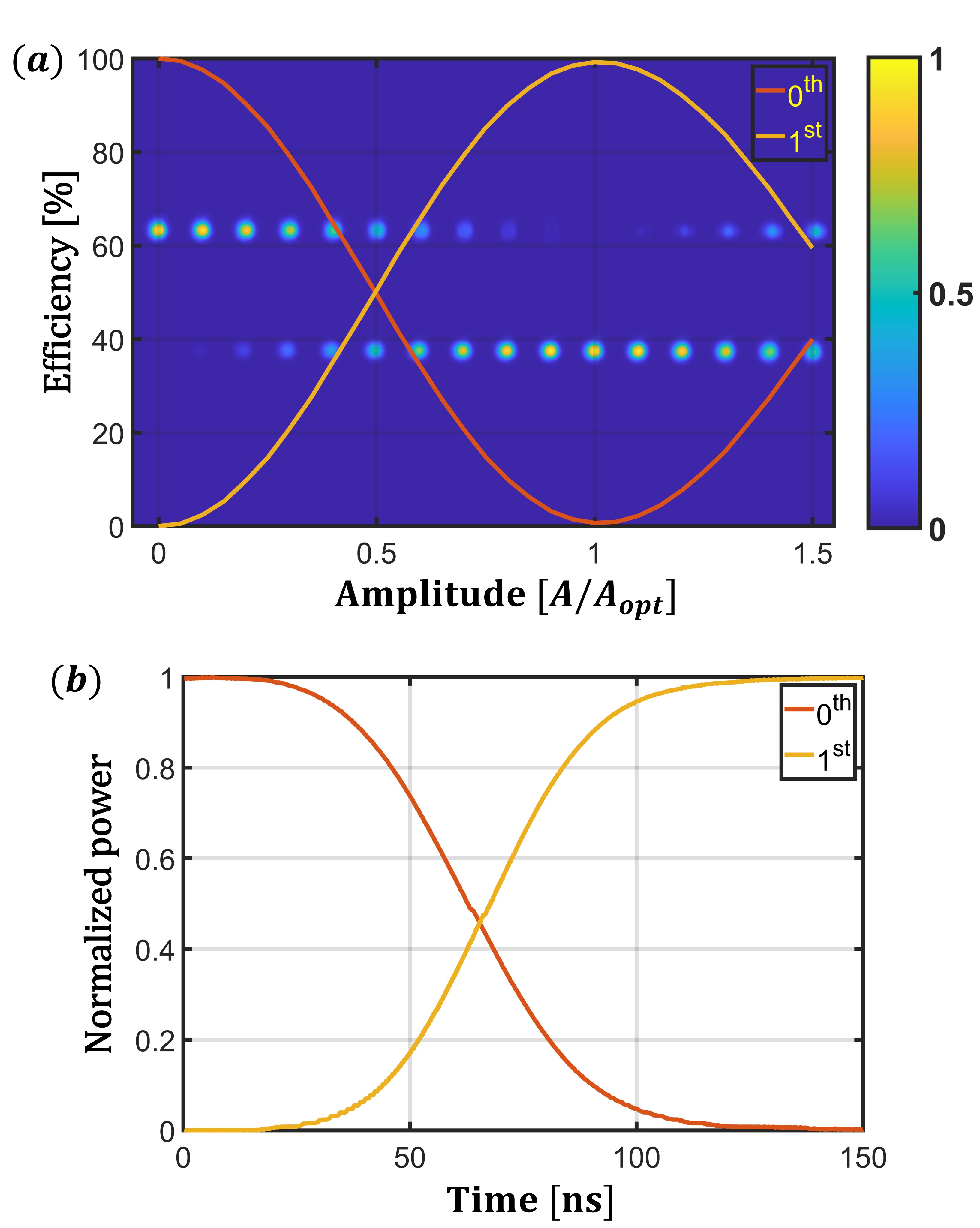}
    \caption{CP-AOM as full contrast optical router. (a): The $1^{\rm st}$-order diffraction efficiency $R$ and $0^{\rm th}$-order transmission efficiency $T$ (excluding insertion loss) as a function of RF modulation amplitude $A$, evaluated from the intensity profile data as those in Fig.~\ref{fig:img}. Sixteen representative intensity profiles are selected as background, with the respective RF-amplitude $A$ aligned to the x-axis. (b): The $0^{\rm th}$ and $1^{\rm st}$-order output power during a nanosecond switch-on of CP-AOM. 
    }
    \label{fig:RT} 
     \end{figure}
We denote the RF-modulation amplitudes for the optimal CP-AOM diffraction in Fig.~\ref{fig:img}c as $A_{1,2}=a_{1,2} A_{\rm opt}$, with $a_{1,2}\approx 1$ to be alignment-dependent relative amplitudes. As demonstrated in Fig.~\ref{fig:RT}a, for $A_{1,2}= a_{1,2} A$ and by varying $A$ between $0$ and $A_{\rm opt}$ (thereby with $K$ adjusted between 0 and $\pi/2L$, Appendix~\ref{sec:hLoss}), arbitrary division of optical power between the $0^{\rm th}$ and $1^{\rm st}$ diffraction orders can be achieved. In particular, by switching on/off CP-AOM, an incident beam is routed to the $1^{\rm st}$/0$^{\rm th}$ output diffraction order on-demand. While not shown, we note that when using CP-AOM, the optical phase for the $1^{\rm st}$ diffraction order output can be adjusted electronically with $\bar\varphi=(\varphi_1+\varphi_2)/2$, a task that is difficult to realize with EOM-based fast optical routers~\cite{Rudolph2017, Madsen2022, Yu2023}. 

To test the CP-AOM control speed, we simultaneously switch on the RF-signals driving AOM$_{1,2}$ (Fig.~\ref{fig:AOM}a), and to measure the fiber-coupled $I_{0,1}$ outputs with the  PDA8GS detector. To ensure synchronized switching of AOM$_{1,2}$, the time-of-arrival for the two TTL signals driving the CMOS switches (Analog Device ADG904 module) are relatively adjusted. As shown in Fig.~\ref{fig:RT}b, with the switch-on the $I_1$ signal rises up from zero to the value associated with the $\sim 99\%$ efficiency, while the $I_0$ signal effectively falls to zero. Similar behavior is observed during the switch-off. A 10$\%$-to-90$\%$ transient time of $\tau_s\approx 80$~ns is found in all cases. The transient time is consistent with the $w=185~\mu$m beam size, and is similar to those achieved with the less-efficient single-AOM operations. 

We note optical routing with $\tau_s \sim 80~$ns switching time is already useful for replacing EOMs to manipulate quantum light in various experiments~\cite{Rudolph2017, Madsen2022, Yu2023}. Even faster CP-AOM control can be realized by using higher-frequency $\omega_s$ with a more focused laser beam, with a potential switching time down to 10~nanosecond level.  On the other hand, when the requirement is to periodically rout a periodic pulse train into two directions, then CP-AOM in counter-propagating geometry (Sec.~\ref{sec:geometry}) can be exploited for achieving GHz-level switching rate~\cite{Liu2022c}.

\section{Discussions}\label{sec:conclusion}
The CP-AOM configuration in this work exploits high precision 4-F imaging optics to facilitate coherent double-diffraction by a pair of AOMs. The idea of optical imaging can be extended further to CP-AOM with $N=1$, single-AOM and $N>2$ multiple AOMs. The possibilities are discussed in the following. 

\subsection{Single AOM implementation}\label{sec:one}
In the Fig.~\ref{fig:AOM}a setup, a mirror can be inserted in the middle plane of the 4-F system. Then, without AOM$_2$, the retro-reflection supports operation of CP-AOM with AOM$_1$ alone. We refer this way of operating CP-AOM as a Michelson-interferometer configuration. In this configuration, the relative phase $\phi_{12}$ between the double diffraction is time-dependent, $\phi_{12}(t)=2\omega_s t+\phi_{12}(0)$, similar to CP-AOM in the counter-propagating geometry~\cite{Liu2022c} (Sec.~\ref{sec:geometry}). Clearly, in the single-AOM operation half of the CP-AOM control degrees of freedom are lost. Nevertheless, as being discussed in ref.~\cite{Liu2022c}, the Michelson-CP-AOM can be particularly useful for controlling the repetition rate of mode-locked lasers.

Alternatively, one may replace AOM$_2$ with a retro-reflection mirror around the AOM$_2$-location where the diffraction orders from AOM$_1$ cross each other. The orders are reversed by the mirror reflection. Similar to the co-propagating CP-AOM explored in this work, the double-diffracted output orders are phase-stable. We refer this way of operating CP-AOM as a Sagnac-interferometer configuration. In this configuration, the relative phase $\phi_{12}$ is adjusted by the optical delay between the two acoustic diffractions, $\phi_{12}\approx 8\omega_s f/c$, which can be optimized by varying the distance between the two imaging lenses ($l_{1,2}$ in Fig.~\ref{fig:img}a). Similar to the Michelson-CP-AOM, in the Sagnac-CP-AOM the number of AOM-control parameters are halved. Since $\phi_{12}$ cannot be electronically adjusted, the wideband operation as in Fig.~\ref{fig:scan} is not supported by the single-AOM operation.

We have constructed the single-AOM-based CP-AOM in the Michelson and Sagnac configurations to control the mode-locked~\cite{Ma2020} and CW lasers~\cite{Ji2023} in our lab. Depending on the type of AOM and operating $\omega_s$, optimal diffraction efficiencies of $R>97\%$ are regularly achieved.


\subsection{Beyond 2-AOM}\label{sec:N}
The CP-AOM with $N=2$ as in Fig.~\ref{fig:AOM}a can be expanded to $N>2$ using additional $4-F$ imaging links. The $N>2$ CP-AOM may support even more precise diffraction and potentially more optical control functionalities. Within the 2-mode approximation (Sec.~\ref{sec:mode}), the possibility of enhancing the first-order Bragg diffraction with $N>2$ has been discussed in ref.~\cite{Liu2022c}. Beyond the analysis there, we note CP-AOM with $N>2$ include additional $\delta L_j, j=3,...,N$ translational degrees of freedom, similar to that is outlined in Fig.~\ref{fig:AOM}a, to further improve the CP-AOM performance. Furthermore, precise optical wavefront engineering appears feasible to enable efficient high-order Bragg diffraction or to generate symmetric output orders~\cite{Wu2005, Hughes2007}. The control degrees of freedom offered by the additional AOMs are useful resources to ensure multi-color beams controlled in a manner that is either sensitive or insensitive to the laser wavelength for {\it e.g.}, spatial-temporal shaping of ultrafast pulses~\cite{Chen2022c}.

Clearly, since the overall insertion loss in CP-AOM increases with $N$, a low insertion loss per AOM is particularly important for achieving the versatile CP-AOM functionality. To this end, we note that the $N-$AOM diffraction requires $\theta=\pi/N$ rotation (Eq.~\eqref{eq:Bloch}) for achieving nearly ideal first-order Bragg diffraction. Therefore, increasing $N$ drastically reduces~\cite{Herold2012} the RF-power requirement, $P_{\rm RF}\propto 1/N^2$ for each daughter-AOM, for efficient acousto-optical modulation. This technical advantage may help wider application of quartz-based AOMs~\cite{Vidne2003} that are typically more difficult to drive but support higher laser power and lower insertion loss.

\subsection{Summary and outlook}
Light sometime needs to be modulated to be useful. Acoustic-optical modulation, with the speed and accuracy, is a uniquely important technique for advancing optical science at frontiers. Inspired by advances of composite-pulse quantum control~\cite{Wu2005,Low2016},  we demonstrate that AOM efficiency can be perfected by composite diffraction~\cite{Liu2022c} in a resource-efficient manner without sacrificing the control bandwidth. This, combined with the versatile functionalities of CP-AOM (Sec.~\ref{sec:geometry})~\cite{Liu2022c} and the substantially reduced RF-power requirement (Sec.~\ref{sec:N}), may help to greatly expand our ability to accurately control light with low-loss acousto-optics.

Experimentally, the $R=99.2\%$ diffraction efficiency achieved in this work with the $N=2$ CP-AOM is likely limited by wavefront aberrations. More precise imaging optics would help to improve the efficiency further to the $R=99.9\%$ level, thereby also enhancing the zeroth-order rejection ratio to beyond 40~dB (Fig.~\ref{fig:scan}b). With improved optics and assuming $1\%$ level insertion loss unrelated to diffractions, iterative application of CP-AOM would support $\Delta \omega = N_i \omega_s$ with $N_i \sim 100$, thereby pushing the free-space acousto-optical frequency-shifting range deeply into the microwave regime~\cite{Zhou2020}. The low insertion loss combined with nearly ideal diffraction should also facilitate the application of CP-AOM as a fast two-port optical router for quantum light~\cite{Rudolph2017, Madsen2022, Yu2023}, or for coherently splitting/combining pulsed and CW lasers~\cite{Liu2022c, Hansel2017, Wang2021}.  Similar to composite pulse quantum control~\cite{Low2016}, the CP-AOM functionalities expand in $N-$AOM systems with large $N$. Examples include achieving high-order Bragg diffraction~\cite{Wu2005,Hughes2007}, general wavefront engineering~\cite{Jabbari2024}, and spatial-temporal control of ultrafast pulses~\cite{Chen2022c}.

The CP-AOM technique discussed in this work is in direct analogy to composite pulse control of matterwave~\cite{Wu2005,Hughes2007}. Nevertheless, a distinct feature in CP-AOM is the precise 4-F imaging for the effective wavefront reversal. Aided by developments of matterwave lensing techniques~\cite{Kovachy2015,Asteria2021}, we also expect the momentum echo and high-order rephasing effects discussed in this work to be exploited in atom interferometry, to enhance the matterwave control efficiency/bandwidth~\cite{Kirsten-Siemb2023, Louie2023}.

\section*{Acknowledgements}
This work is supported by the National Key Research Program of China (Grant No. 2022YFA1404204), the National Natural Science Foundation of China (Grant No. 12074083), the Natural Science Foundation of  Shanghai (Grant No. 23dz2260100), and the Shanghai Science and Technology Innovation Action Plan (Grant No. 24LZ1400300).

\section*{Disclosures}
The authors declare no conflicts of interest.

\appendix
\section{Derivation of Eqs.~\eqref{eq:R2level}\eqref{eq:R2levelb}}\label{sec:derive}
We consider Eq.~\eqref{eq:UAOM} under 2-mode approximation, and rewrite $H=H_0(k_\perp)+V(K,\phi)$ in terms of Pauli matrices,
\begin{equation}
H=\frac{1}{2}(\Delta \sigma_z- K \sigma_\phi).
\end{equation}
Here $\sigma_x=(0, 1;1, 0)$, $\sigma_y=(0, -i;i, 0)$, $\sigma_z=(1, 0;0, -1)$, so that $\sigma_\phi=(0 , e^{i\phi};e^{-i\phi}, 0)$.

Single-AOM propagator according to Eq.~\eqref{eq:UAOM} is:
\begin{equation}
U={\rm cos}(\frac{\tilde \theta}{2})-\frac{i\varepsilon}{\sqrt{1+\varepsilon^2}}{\rm sin}(\frac{\tilde \theta}{2}) \sigma_z-\frac{i}{\sqrt{1+\varepsilon^2}}{\rm sin}(\frac{\tilde \theta}{2})\sigma_{\phi}\label{eq:U2}
\end{equation}
Here, for the collimated laser beam input, we introduce the small parameter $\varepsilon=\Delta/K$. For $K=\theta/L$, we further have $\varepsilon=\frac{k_s}{ k}\frac{k_\perp L}{\theta}$. The actual rotation angle is accordingly expressed as $\tilde\theta=\sqrt{1+\varepsilon^2}\theta$.

With $L'=(L+\delta L)/2$, the free propagator $U_0$ is simply given by
\begin{equation}
    U_0= e^{-i \varepsilon K L' \sigma_z}  \label{eq:U0}
\end{equation}

With $C_n(k_\perp)=\delta_{n,0}$ incidence, we focus on $R(k_\perp)=|U_{0 1}|^2$ with $U=U_1$ (single AOM) or $U=U_c$ (CP-AOM), as in the main text. For single AOM, it is easy to show that with $\theta=\pi$
\begin{equation}
R(k_\perp)=1-(\frac{k_s}{\pi k})^2(k_\perp L)^2+\mathcal{O}(\varepsilon^4),
\end{equation}
which leads to $a_0=(k_s/\pi k)^2$ in Eq.~\eqref{eq:R2level}.

Next, for $U_c=U_2 U_0^{-1}U_1$, we find
\begin{equation}
\begin{aligned}
&R(k_\perp)=\frac{4\sin^2(\frac{\tilde{\theta}}{2})}{1+\varepsilon^2}\times \\
    &\left| {\rm cos} (\varepsilon K L') \cos(\frac{\tilde{\theta}}{2})+ {\rm sin} ( \varepsilon K L')\frac{\varepsilon}{\sqrt{1+\varepsilon^2}}\sin(\frac{\tilde{\theta}}{2})
    \right|^2    \label{eq:Rperp}
    \end{aligned}
\end{equation}

We now consider $\theta=\pi/2$~\cite{Liu2022c}, let $\delta =\delta L/L$, and expand Eq.~\eqref{eq:Rperp} to the fourth order in $\varepsilon$ to obtain Eq.~\eqref{eq:R2levelb} in the main text, with
\begin{equation}
\begin{aligned}
a(\delta L)&=\left(\frac{\pi\delta+\pi-4}{2}\right)^2 \left(\frac{k_s}{\pi k}\right)^2,\\
b(\delta L)&=\Big(\frac{\pi^4}{48}(1+\delta)^4-\frac{\pi^3}{3}(1+\delta)^3+\\
&(3+6\delta +2\delta^2)\pi^2-12(1+\delta)\pi+16\Big)\left(\frac{k_s}{\pi k}\right)^4.  \label{eq:ab}
\end{aligned} 
\end{equation}

By setting $a(\delta L)=0$, we find
\begin{equation}
\delta L_{\rm opt}=(\frac{4}{\pi}-1)L \label{eq:dLTaylor}
\end{equation}
leading to $b=(4-\pi)^2 (k_s/\pi k)^4$ in Eq.~\eqref{eq:R2levelb}.




\section{Numerical integration}\label{app:num}

The diffraction dynamics in CP-AOM is characterized by the $U_c$ propagator in Eq.~\eqref{eq:Uc}. Numerically, $U_c$ is obtained by integrating either Eq.~\eqref{eq:2D_SEq} or Eq.~\eqref{eq:Norder Raman-Nath}. The free-space propagator $U_0$ is obtained by setting $K=0$ in these equations. To investigate the diffraction dynamics for a specific $k_\perp$ component, we choose to integrate Eq.~\eqref{eq:Norder Raman-Nath} following proper mode truncation (Sec.~\ref{sec:mode}). For example, the Fig.~\ref{fig:AOM}(b,d) illustrations are with $n_0=1$ to investigate the $n=0,1$ two-mode dynamics. 

For a Gaussian beam input with continuous $k_\perp$ distribution, such as for Fig.~\ref{fig:AOM}(c), we choose to  directly integrate Eq.~\eqref{eq:2D_SEq} with the incident Gaussian $\mathcal{E}(x,0)$. With the $\partial_t$ term ignored, Eq.~\eqref{eq:2D_SEq} is rewritten as:
\begin{equation}
i\partial_z\mathcal{E}(x,z) = (H_0 + V(K,\phi))\mathcal{E}(x,z).\label{eq:2D_SEqb}
\end{equation}
Here, $V(K,\phi)=-(\frac{K}{2}e^{i (k_s x+\phi)}+c.c.)$ and $H_0 = -\frac{1}{2k} \partial^2_x$ are equivalent to those in Eq.~\eqref{eq:Norder Raman-Nath}, but are expressed in the real $x-$space. 
Numerical integration of Eq.~\eqref{eq:2D_SEqb} is carried out using a split-operator method~\cite{Suzuki1976}, with small $z$-steps. The method alternates between diagonalizing the the $V(K,\phi)$ term in real space and $H_0$ term in reciprocal space. To evaluate $R$ in Eq.~\eqref{eq:R} for the Gaussian beam, according to Eq.~\eqref{eq:epsilon_k0}, and more generally $R_n\equiv \int {\rm d}k_\perp R_n(k_\perp) f(k_\perp)$ for the $n^{\rm th}-$order output, we simply integrate the the intensity of the output beam in the reciprocal $k_x-$space within $|k_x - n k_s| < k_s /2$.




\section{High-order loss during partial splitting}\label{sec:hLoss}

\begin{figure}[htbp]
    \centering
    \includegraphics[width=1\linewidth]{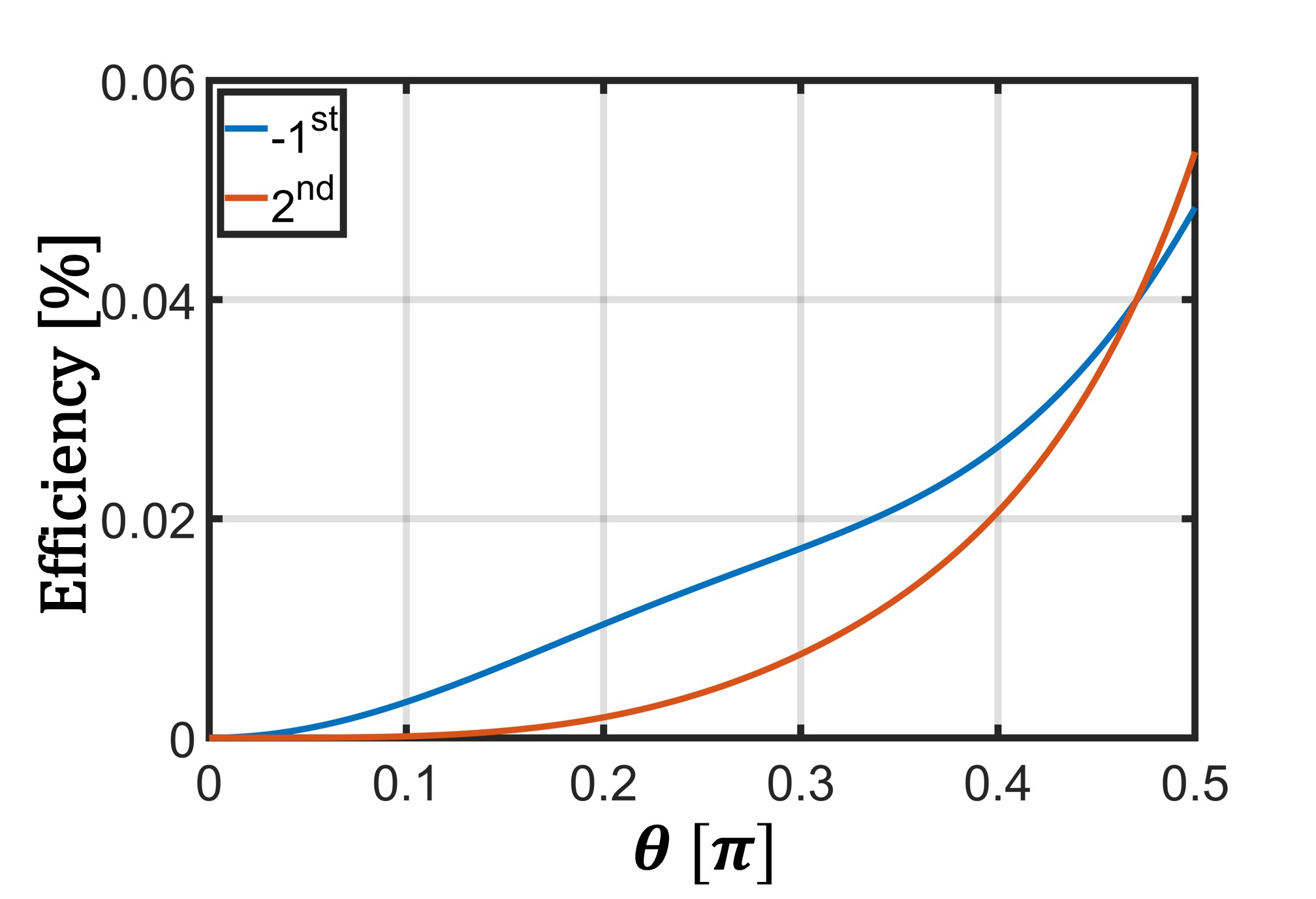}
    \caption{The $n=-1,2$ high-order loss as a function of $\theta$ for the Fig.~\ref{fig:AOM}a setup, at the optimal $\delta L_{\rm opt}\approx 0.27 L$ (Eq.~\eqref{eq:dLTaylor}) that minimizes the losses at $\theta=\pi/2$ according to Eqs.~\eqref{eq:R2levelb}~\eqref{eq:hLossC}, and Fig.~\ref{fig:AOM}c. 
    }
    \label{fig:hLoss} 
     \end{figure}

In Sec.~\ref{sec:highOrder} we discussed optimal choices of $\delta L$ to suppress high-order diffraction losses. As by Eq.~\eqref{eq:hLossC}, the optimal $\delta L$s that minimize $\mathcal{L}_c$ loss depend on the complex coefficient $\mathcal{A}_{1,2}$, which are decided by AOM diffraction dynamics to be sensitive to $K$. A natural question arises, as whether the $\delta L$ optimized at $K=\pi/2L$ remains efficient to suppress the high-order losses for varying $K$. A general analysis of $\mathcal{L}_c(K;\delta L_{\rm opt})$ appears quite complicated. Instead, in Fig.~\ref{fig:hLoss} we provide a numerical example of the $\mathcal{L}_c$ vs $K=\theta/2L$. Here $\delta L=\delta L_{\rm opt}$ according to Eq.~\eqref{eq:dLTaylor} is optimized at $\theta=\pi/2$ as in Fig.~\ref{fig:AOM}c. We then numerically study the high-order loss as $\theta$ varies between $0$ and $\pi/2$. The full-order simulation shares the same parameters as those for Fig.~\ref{fig:AOM}c. We find that the high-order losses decrease monotonically with $K$. Similar to Fig.~\ref{fig:AOM}c, we verified  this observation to be stable when the acoustic field amplitude $\eta(z)$ modifies perturbatively in shape. Therefore, we expect the impact of high-order losses  to be accordingly limited when using CP-AOM as a programmable beamsplitter, as experimentally demonstrated with Fig.~\ref{fig:RT} in the main text.

~

~

~

~

~

~

\bibliography{HEAOM}
\bibliographystyle{revTex2}

\end{document}